\documentclass[12pt]{iopart}
\usepackage{amssymb}
\usepackage{iopams,bm,verbatim}
\usepackage[T1]{fontenc}
\def\Journal#1#2#3#4{{#4} {\it #1} {\bf #2}, #3 }

\def\Xc{\bar{X}}
\def\td{\tau'}
\def\tdbar{\overline{\tau}'}
\def\tbar{\overline{\tau}}
\def\rd{\rho'}

\def\phidbar{\overline{\phi}'}
\def\phibar{\overline{\phi}}
\newcommand{\w}[1]{\bm{#1}} 

\def\tho{\textrm{\TH}}
\def\thd{\tho '}
\def\et{\eth}
\def\etd{\eth '}
\newcommand{\be}{\begin{equation}}
\newcommand{\ee}{\end{equation}}

\begin{document}

\title{Two special classes of space-times admitting a non-null valence two Killing spinor}

\author{N.~Van den Bergh}

\address{Ghent University, Department of Mathematical Analysis IW16, \\ Galglaan 2, 9000 Ghent, Belgium}
\eads{\mailto{norbert.vandenbergh@ugent.be}}

\begin{abstract}
    Non-conformally flat space-times admitting a non-null Killing spinor of valence two
are investigated in the Geroch-Held-Penrose formalism. Contrary to popular belief these space-times are not all explicitly known.
It is shown that the standard construction hinges on the tacit assumption that certain integrability conditions hold, implying two algebraic relations, $KS_1$ and $KS_2$, for the spin coefficients and the components of the Ricci spinor. An exhaustive list of (conformal classes of) space-times, in which either $KS_1$ or $KS_2$ are violated, is presented. The resulting space-times are each other's Sachs transforms, in general admit no Killing vectors and are characterized by a single arbitrary function.
\end{abstract}

\pacs{04.20.Jb}

\section{Introduction}
The concept of a Killing spinor has its origins in the work of Walker and Penrose~\cite{WalkerPenrose70}
who, in the class $\mathcal{D}_0$ of Petrov type D solutions of Einstein's vacuum field
equations with cosmological constant, demonstrated the existence of a valence two symmetric spinor
$X_{AB}$, satisfying the conformally invariant (twistor) equation
\begin{equation}\label{eqKS}
\nabla _{A'(A}X_{BC)}=0 .
\end{equation}
The original significance of such a spinor is that its existence in a space-time ($\mathcal{M}$, g)
determines a constant of the motion along any null geodesic. The existence of this constant
of the motion in $\mathcal{D}_0$ may be equivalently derived from the separability of the Hamilton-Jacobi equation for the null geodesics therein \cite{CzaporMcLen1982}.
Introducing the two-form
$D_{ab}= X_{A B} \epsilon_{A'B'}+\Xc_{A'B'} \epsilon_{AB}$,
called a conformal Killing-Yano (CKY) tensor, the equation corresponding to (\ref{eqKS}) reads~\cite{Floyd73,Tachibana69}
\begin{equation}\label{eqKY}
\nabla_{(a} D_{b)c}= \frac{1}{3} \nabla_d (g_{ab} {D^d}_c-{D^d}_{(a}g_{b)c}),
\end{equation}
or, equivalently~\cite{Glass},
\begin{equation}
 3 \nabla_c D_{ab} = 3 \nabla_{[c} D_{ab]}-2 \nabla_d {D^d}_{[a}g_{b]c} .
\end{equation}

As a consequence these space-times are also frequently called CKY space-times~\cite{FerrandoSaez2005}. When the right hand side of (\ref{eqKY}) vanishes,
 $D_{ab}$ is called a Killing-Yano tensor
 or a Penrose-Floyd tensor~\cite{Papacostas1985}. This happens if and only if $X_{AB}$ satisfies the additional skew
Hermiticity condition
\begin{equation}\label{eqreal}
{\nabla^B}_{A'}X_{AB}+{\nabla_A}^{B'}\Xc_{A'B'}=0.
\end{equation}
A space-time admitting a valence two Killing-Yano tensor
will be called a KY space-time: all these have been determined explicitly \cite{DietzRudiger1, DietzRudiger2,KamranMcL1984, Papacostas1985}. Note that a CKY space-time is \emph{not necessarily} conformally related to a KY space-time, as examples in \cite{Jeffryes1984} and \cite{McLenVdB1993} show\footnote{to add to the confusion spinors satisfying (\ref{eqKS}) were termed conformal Killing spinors in \cite{CarterMcLen79} and \cite{CzaporMcLen1982}, while the name Killing spinor or Killing spin two-forms was reserved for spinors or their tensor analogues satifying also (\ref{eqreal}). As explained in \cite{Jeffryes1984} it is better to refrain from this terminology.}.
The square $K_{ab}={D_a}^c D_{cb}$ of a KY tensor $D$ is a Killing tensor (KT), namely a symmetric tensor satisfying the equation
\begin{equation}\label{KT}
\nabla_{(a} K_{bc)}=0
\end{equation}
and giving rise to a quadratic first integral of the geodesic equation. Similarly the square of a CKY tensor is a conformal Killing tensor (CKT), namely a symmetric tensor satisfying
\begin{equation}\label{CKT}
\nabla_{(a} K_{bc)}= g_{(ab} k_{c)},
\end{equation}
where $k_c$ is obtained by contraction of (\ref{CKT}) with $g^{ab}$. Conformal Killing tensors are of interest inasmuch as they give rise
to quadratic first integrals for null geodesics. For the sequel it is worth noting that, if $K_{ab}$ is a conformal Killing tensor, then so is $K_{ab}+f g_{ab}$ with $f$ an arbitrary function (both defining the
same first integrals of the null geodesic equation). This implies that one can always assume $K_{ab}$ to be traceless, implying $k_c=\frac{1}{3}\nabla_m {K^m}_c$.
Note that the converse of the property above not necessarily holds: not every (conformal) Killing tensor is the square of a (conformal) Killing-Yano tensor and conditions which a (C)KT space-time has to obey in order to be a (C)KY space-time have been discussed in \cite{Collinson1976} and \cite{FerrandoSaez2005}. These conditions are satisfied for example in those subclasses of the type D vacuum solutions and the aligned non-null Einstein-Maxwell solutions that admit a Killing tensor~\cite{HughstonSommers1973,Collinson1976, Stephani1978}.
In \cite{CollFerrandoSaez2006, FerrandoSaez2007} it was furthermore shown that every type D CKT space-time with a Killing tensor of Segre type [(11),(11)] is CKY.
\\

The Weyl spinor of a non-conformally flat space-time admitting a \emph{non-null} valence two Killing spinor $X_{AB}$ is necessarily of Petrov type D with repeated principal spinors aligned with the principal spinors of $X_{AB}$ \cite{PlebHacyan76}, while the Weyl principal null
directions define geodesic shear-free\footnote{this property also holds~\cite{DietzRudiger1980, HauserMalhiot1976} for the two null eigendirections of any (conformal) Killing tensor of Segre type [(11),(11)]).} null congruences. Henceforth I will call these KS space-times, to distinguish them from the more general CKY space-times, for which the Petrov type also can be N or O. Obviously, as (\ref{eqKS}) is conformally
invariant, KS space-times can only be determined up to an arbitrary conformal transformation
of the metric.
KS space-times necessarily include all those which are conformally related to the Petrov type D KY space-times. The inclusion is strict: examples exist which satisfy (\ref{eqKS}) but not (\ref{eqreal}): they appear in the classes $\mathcal{D}$ and $\mathcal{D}_0$ of solutions of Einstein's electrovac and vacuum
field equations with cosmological constant for a non-null aligned Maxwell field as the
Kinnersley Case III solutions \cite{CzaporMcLen1982,DebeverMcLen1981}.
A KS space-time always admits a conformal representant in which the trace of the associated conformal Killing tensor is constant. In this representant the conformal Killing tensor becomes a Killing tensor, but, by construction, has two constant eigenvalues (the reason why this is so will become clear in section 2). As the non-constancy of the eigenvalues plays an essential role in the construction of appropriate coordinates~\cite{HauserMalhiot1976,Jeffryes1984, Papacostas1983a, Papacostas1983b} and in the ensuing separability properties, one may ask whether a conformal transformation exists which preserves the existence of the Killing tensor and which leads to non-constant eigenvalues (the so-called non-singular Killing tensors~\cite{HauserMalhiot1978}). In \cite{Jeffryes1984} the answer to this question was taken to be affirmative,
as a consequence of the tacit assumption of two integrability conditions, introduced in section 2 below as $KS_1$ and $KS_2$.
The resulting list of canonical line-elements of KS space-times therefore turned out to be incomplete. This was partially remedied in \cite{McLenVdB1993}, where the extra line-elements were constructed which arise in the `I-N family' of \cite{Jeffryes1984}, defined by the vanishing of one of the spin coefficients $\rho$ or $\mu$. In the present paper the extra line-elements are constructed which arise in the general `family I', defined by the non-vanishing of $\rho$, $\mu$, $\pi$ and $\tau$, and by assuming that only one of the two integrability conditions $KS_1$ or $KS_2$ holds. In order to fully exploit the symmetry provided by the alignment of the principal null directions of the Weyl and Killing spinor, the Geroch-Held-Penrose formalism~\cite{GHP} is used throughout: the $KS_1$ and $KS_2$ families are then each other's Sachs transforms~\cite{GHP}.

\section{Preliminaries}
Writing the Killing spinor as $X_{AB}=X o_{(A} \iota_{B)}$ and using $o, \iota$ as the basis spinors for the GHP formalism, the components of (\ref{eqKS}) imply (see \cite{Jeffryes1984} for details)
\begin{equation}\label{eqKSa}
\kappa=\sigma=0
\end{equation}
and
\begin{eqnarray}
\tho X = -\rho X, \label{KSb1}\\
\et X = -\tau X, \label{KSb2}
\end{eqnarray}
together with their `primed versions' and $X'=X$, namely $\kappa'=\sigma'=0$ and $\thd X =-\rd X$, $\etd X = -\td X$.
>From (\ref{eqKSa}), the application of the $[\et,\, \tho]$ commutator to $X$ and the GHP equations one finds
that the principal null directions of the Killing spinor and the Weyl spinor are aligned, i.e.~$\Psi_2$ is the only non-vanishing component of the Weyl spinor.
The equations (\ref{eqKSa},\ref{KSb1},\ref{KSb2}) are invariant under conformal transformations $g\rightarrow\Omega^2 g$, $X\rightarrow\Omega X$ and a conformal representant $(\mathcal{M},\hat{g})$ can be fixed by imposing $|\hat{X}|=1$. In the manifold $(\mathcal{M},\hat{g})$ we have then
\be\label{memoke1}
\hat{\rho}+\overline{\hat{\rho}}=\hat{\rho}'+\overline{\hat{\rho}}'=\hat{\tau}+\overline{\hat{\tau}}'=0,
\ee
while in the $(\mathcal{M},g)$ manifold with $g=
\Omega^2\hat{g}$ one has $\Omega^2=X\overline{X}$.
The cases $\tau=\td =0$ and $\rho$ or $\rd=0$ have been discussed in \cite{Jeffryes1984} and \cite{McLenVdB1993}, so henceforth $\rho,\rd,\tau,\td$ will be taken $\neq 0$.

Defining the trace-free tensor $P_{ab}= X_{AB}\overline{X}_{A'B'}$, or, with null vectors $\ell_a = o_Ao_{A'}$, $n_a=\iota_A \iota_{A'}$ and $m_a=o_A\overline{\iota}_{A'}$,
\be
P_{ab}=\frac{\Omega^2}{2}(m_{(a} \overline{m}_{b)}+\ell_{(a}n_{b)}),
\ee
one sees that $P_{ab}$ obeys the CKT equation (\ref{CKT}). This implies that $P_{ab}$ will be the trace-free part of a Killing tensor $K_{ab}$, \emph{provided} a solution $K_{ab}=P_{ab}+\frac{1}{4} K g_{ab}$ exists of the KT equation (\ref{KT}).  Contraction implies the necessary and sufficient condition
\be
\nabla_a {P^a}_c +\frac{3}{4} \nabla_c K =0.
\ee
In components\footnote{there is a sign difference with \cite{Jeffryes1984}, which is probably due to the use of a
different signature convention; I use $\ell^a n_a=1 =-m^a \overline{m}_a$. The correspondence with the Newman-Penrose operators and the basis one-forms
is taken as $(m^a,\overline{m}^a,n^a, \ell^a) \leftrightarrow (\delta, \overline{\delta}, \Delta, D)$ and $(-m_a,-\overline{m}_a,\ell_a,n_a) \leftrightarrow
(\w{\omega}^1,\w{\omega}^2,\w{\omega}^3,\w{\omega}^4)$} this becomes
\begin{eqnarray}
\tho K -\Omega^2(\rho+\overline{\rho})=0, \label{KTcompsa} \\
\et K +\Omega^2(\tau+\tdbar) = 0 \label{KTcompsb}.
\end{eqnarray}
or, 
in terms of the eigenvalues $a=(\Omega^2-K)/4$ and $b=(\Omega^2+K)/4$ of
\be
K_{cd}= 2 a m_{(c}\overline{m}_{d)}+2 b \ell_{(c} n_{d)} :
\ee
\begin{eqnarray}
\tho b =0 \label{eqthb}, \\
\et a=0 \label{eqeta}.
\end{eqnarray}
By (\ref{KSb1},\ref{KSb2}) and $\Omega^2=X \overline{X}$ this also implies
\begin{eqnarray}
\tho a = -(a+b) (\rho+\overline{\rho}), \label{KTextra}\\
\et b = -(a+b) (\td +\tbar), \label{KTextrb}
\end{eqnarray}
such that (\ref{eqthb},\ref{eqeta},\ref{KTextra},\ref{KTextrb}) alternatively can be written as
\begin{eqnarray}
\mathrm{d} a = -(a+b) [(\rho+\overline{\rho})\w{\omega}^4+(\rho'+\overline{\rho}')\w{\omega}^3] \label{dmaina} \\
\mathrm{d} b = -(a+b)[(\tau+\tdbar)\w{\omega}^1 + (\overline{\tau}+\td)\w{\omega}^2] \label{dmainb},
\end{eqnarray}
which exhibits clearly the relevance of the existence of a Killing tensor with non-constant eigenvalues for the construction of appropriate coordinates.
It also becomes obvious now, as mentioned in the introduction, that all KS space-times admit at least one conformal representant
(namely $(\mathcal{M},\hat{g})$ or its constant re-scalings) in which a Killing tensor exists, but that by (\ref{memoke1}) this
particular Killing tensor has constant eigenvalues.
\emph{If one assumes} that there is a non-trivial conformal representant in which $a$ and $b$ are \emph{not} both constants, i.e.~if one assumes that non-constant solutions exist of the system (\ref{eqthb},\ref{eqeta}) or (\ref{dmaina},\ref{dmainb})
then extra integrability conditions result (namely $\mathrm{d}\mathrm{d} a= \mathrm{d}\mathrm{d} b=0$). It is preferable to study these equations in the $(\mathcal{M},\hat{g})$ manifold, where the $\hat{}$ symbol from here onwards will be dropped: the remaining spin coefficients are then $\rho, \rd$ and $\tau=-\tdbar$.  Under the conditions (\ref{memoke1}) the integrability conditions for the system (\ref{KSb1},\ref{KSb2}) simplify to
\begin{eqnarray}\label{specKS}
\thd \rho -\tho \rd =0 , \label{specKSa}\\
\et \td -\etd \tau =0, \label{specKSb} \\
\tho \td -\etd \rho = 0 .\label{specKSc}
\end{eqnarray}
The GHP equations, together with the directional derivatives of (\ref{memoke1}), reduce to the system
\begin{eqnarray}
\tho \rho =0, \label{thrho}\\
\et \rho = 2 \rho \tau +\Phi_{01},\label{etrho}
\end{eqnarray}
\begin{eqnarray}
\tho \tau = 2 \rho \tau + \Phi_{01}, \label{thtau} \\
\et \tau = 0, \label{ettau}
\end{eqnarray}
\be
\tho \rd -\et \td
= -\rho \rd -\tau \tbar-\Psi_2-\frac{1}{12} R \label{thrd_ettd}\\
\ee
and impose the following restrictions on the curvature:
\be
\Phi_{00}=-\rho^2,
 \Phi_{02}=-\tau^2.
\ee
All these equations
are to be considered as being accompanied by their complex conjugates as well as their `primed versions' (taking into account $\Phi_{00}'=\Phi_{22}$, $\Phi_{02}'=\Phi_{20}$, $\Phi_{10}'=\Phi_{12}$ and (\ref{memoke1})).
The remaining derivatives of the spin coefficients are related by (\ref{specKSa},\ref{specKSb}), while (\ref{specKSc}) becomes an identity under ($\overline{\ref{etrho}}$,
\ref{thtau}').
Also (\ref{specKSb}) and (\ref{thrd_ettd}) imply that the real part of
\be
\Psi_2=E+i H
\ee
can be written as
\be
E=-\frac{R}{12}-\rho \rd +\tau \td.
\ee
It is advantageous now to introduce 0-weighted quantities $u,v$ (real and with $u'=u$, $v'=v$) and $\phi,\phi'$ (complex) for the remaining curvature components as follows:
\begin{eqnarray}
R=8(u-v)-16\rho \rd, \\
\Phi_{11}=u+v-2 \rho \rd,\\
\Phi_{01}=-3 \rho \tau -2\frac{\rho}{\td} \phi .
\end{eqnarray}
Under the ${}'$ operation this also implies $\Phi_{21}=-3 \rd \td -2\rd  \phi ' / \tau $.

The integrability conditions for the system (\ref{eqthb},\ref{eqeta}) become then
\be
\rd \tho a - \rho \thd a = 0 \label{I1a}
\ee
and
\be
\td \et b -\tau \etd b =0 .\label{I1b}
\ee
Acting respectively with $\et$ and $\tho, \thd$ on (\ref{I1a}) and (\ref{I1b}) and eliminating the second order derivatives by means of the $[{\et^{(}}'{}^{)},\, {\tho^{(}}'{}^{)}]$ commutator relations, results in two more conditions,
\be
\rd \overline{\phi}' \tho a + \rho \phi \thd a =0 \label{I2a}
\ee
and
\be
\td \overline{\phi}\et b +\tau \phi \etd b = 0 .
\label{I2b}
\ee
All further derivatives being identically satisfied, the necessary and sufficient conditions for the existence of a non-constant $b$ or $a$
are now clearly seen to be given by
\begin{eqnarray}
KS_1:  \exists \textrm{ non-constant } b \Longleftrightarrow \phi+\overline{\phi} = \phi'+\phidbar = 0, \label{condKS1}\\
KS_2:  \exists \textrm{ non-constant } a \Longleftrightarrow \phi+\overline{\phi}'=0 . \label{condKS2}
\end{eqnarray}
The corresponding space-times will henceforth be called $KS_1$ or $KS_2$ respectively. While the set $KS_1 \cap KS_2$ has been dealt with in \cite{Jeffryes1984} (both conditions tacitly being assumed to be valid), it is at present not known whether or not $KS_1 \cup KS_2$ yields the full set of KS space-times. The purpose of the present paper is restricted to constructing all space-times in which only one of the two conditions is satisfied.
As a consequence these space-times should belong to the sub-classes with one constant eigenvalue of the Hauser-Malhiot CKT space-times \cite{HauserMalhiot1978}. Actually, as it was proven in \cite{CollFerrandoSaez2006,FerrandoSaez2007} that every type D space-time admitting an aligned conformal Killing tensor of Segre type $[(11),(11)]$ is necessarily CKY, it follows that $(KS_1\setminus KS_2)\cup(KS_2 \setminus KS_1)$ ought to be \emph{exactly} the set of Hauser-Malhiot space-times with one constant eigenvalue. As the metrics found in paragraphs 3 and 4 below contain only one arbitrary function, this raises some doubts about the correctness of the results in \cite{HauserMalhiot1978}.

Introducing 0-weighted extension variables $U$ (real) and $V$ (complex), in accordance with (\ref{specKSa}), by
\begin{eqnarray}
\tho \rd = \thd \rho = -i U , \label{eerste}\\
\thd \phi = \rd( 2\frac{\phi \phi'}{|\tau|^2}+V)+i \frac{U\phi}{\rho},
\end{eqnarray}
the Bianchi identities may be succinctly written in the following form:
\begin{eqnarray}
\tho \phi = \frac{2\rho}{|\tau|^2}(|\tau|^4-|\phi|^2), \label{bi1}\\
\et \phi = -\frac{1}{\td}(2 |\tau|^4+i\phi(U-H)-2\phi^2),\\
\etd \phi = -2\frac{|\phi^2|}{\tau}-\td(2\phi+2\overline{\phi}+\rho \rd -2 v +\frac{i}{2 \rho} \tho H),\\
\tho u =\rho(\phi-\overline{\phi}-3 i H),\\
\et v = \frac{\rho\rd}{\td}(\phi-\phidbar)+i\tau(H+2 U),\\
\et H = 2 i \tau(|\tau|^2+2 u -4\rho \rd)-2 i \frac{\rho \rd}{\td}(2 \phi+2\phidbar +V) .\label{laatste}
\end{eqnarray}
Again these equations are accompanied by their primed versions, taking into account $U'=U,V'=V$ and $H'=H$.
Applying the commutators involving $\thd \rho$ and using (\ref{eerste}) yields two further relations, namely
\begin{eqnarray}
\tho U = 2 i \rho (\rho\rd+2\phi +2\phibar-2 v -2 |\tau|^2), \\
\et U = -2 i \frac{\rho \rd}{\td}(2|\tau|^2+V).\label{bilast}
\end{eqnarray}
 Herewith all `first level' integrability conditions on $\rho,\rd,\tau,\td$ are identically satisfied. The next step is to construct the `second level' integrability conditions by applying the commutators to $\phi, u, v, H$ and $U$.

 When neither of the conditions $KS_1$ nor $KS_2$ hold, carrying out this procedure to the end results in a Pfaffian system for the involved variables, together with a set of algebraic relations among the latter. So far it has not been possible to complete the integrability analysis of this system.
 In the next two paragraphs all space-times will be constructed which belong to $KS_1 \setminus KS_2$ or to $KS_2\setminus KS_1$.

\section{$KS_1\setminus KS_2$}

Consider space-times in which condition $KS_1$ holds, but not $KS_2$. A tedious but straightforward calculation, involving successive derivations of (\ref{condKS1}) and the 'second level' integrability conditions mentioned above, one obtains the following \emph{integrable} system:
\begin{eqnarray}
\tho \rho = 0 ,\nonumber \\
\thd \rho = \frac{2\rho\rd}{|\tau|^2} (\phi+\phi'), \nonumber \\
\et \rho = \frac{\rho}{\td}(|\tau|^2-2 \phi),\label{eerster}
\end{eqnarray}
\begin{eqnarray}
\tho \tau = \frac{\rho}{\td}(|\tau|^2-2 \phi),\nonumber \\
\et \tau = 0 ,\nonumber \\
\etd \tau = \frac{2\rho\rd}{|\tau|^2}(\phi+\phi')+i H ,\label{laatstetau}
\end{eqnarray}
\begin{eqnarray}
\tho \phi = \frac{2\rho}{|\tau|^2}(|\tau|^4+\phi^2),\nonumber \\
\thd \phi = \frac{2\rd}{|\tau|^2}(|\tau|^4-\phi^2+2 \phi\phi'),\nonumber \\
\et \phi = \frac{1}{\td}[ i\phi H-2 |\tau|^4+\frac{2\rho\rd}{|\tau|^2}(\phi^2+\phi\phi')+2 \phi^2 ],\nonumber \\
\etd \phi = \frac{1}{\tau}[ i\phi H-2 |\tau|^4+\frac{2\rho\rd}{|\tau|^2}(\phi^2+\phi\phi')+2\phi^2],
\end{eqnarray}
\begin{eqnarray}
\tho H = \frac{2\rho}{|\tau|^2}[2i\rho \rd(2|\tau|^2+\phi \phi'-\phi^2)+\phi H+4 i |\tau|^4],\label{H1eq}\\
\et H = -8i\tau(|\tau|^2+\rho\rd)-4i\frac{\rho \rd}{\td}(\phi-\phi')-6i\tau E, \label{H2eq}
\end{eqnarray}
\begin{eqnarray}
\tho E = 2i\rho H-8\rho \phi-4\frac{\rho^2\rd}{|\tau|^2}(\phi+\phi'),\label{Eeq1}\\
(\tau\etd-\td \et) E = 4\frac{\rho^2\rd^2}{|\tau|^4}(\phi^2-\phi'^2)+2i\frac{\rho\rd}{|\tau|^2}H(\phi-\phi') . \label{Eeq}
\end{eqnarray}
Introducing 0-weighted real variables $r$ and $m$ by
\be
r^2=Q\rho\rd,\ m=|\tau|\ \ (Q=\pm 1) \label{rmdef}
\ee
and writing $\phi=m^2r^{-2}\phi_0$, $\phi'=m^2r^{-2}\chi_0$, one sees that $\phi_0$, $\chi_0$ are (imaginary) functions of $r$ given by
\be \label{phi0eq}
\frac{\mathrm{d}\phi_0}{\mathrm{d}r}= \frac{\mathrm{d}\chi_0}{\mathrm{d}r} = 2\frac{\phi_0\chi_0+r^4}{r(\phi_0+\chi_0)},
\ee
while (\ref{eerster}-\ref{laatstetau}) yield the exterior derivatives of $r$ and $m$ as
\begin{eqnarray}
\fl \mathrm{d}r = (\phi_0+\chi_0)[\frac{m}{r}(\frac{\tau}{m}\w{\omega}^1-\frac{m}{\tau}\w{\omega}^2)+Q \frac{r}{\rho}\w{\omega}^3+\frac{\rho}{r}\w{\omega}^4],\\
\fl \mathrm{d} m = \frac{i}{2}(2iQ(\phi_0+\chi_0)-H)[\frac{\tau}{m}\w{\omega}^1-\frac{m}{\tau}\w{\omega}^2]+\frac{2m}{r}[Q \frac{r}{\rho}\chi_0 \w{\omega}^3+\frac{\rho}{r}\phi_0 \w{\omega}^4].
\end{eqnarray}
Note that the denominator in (\ref{phi0eq}) is $\neq 0$, as otherwise $\phi_0^2=r^4$ which is in contradiction with $\phi_0$ being imaginary. Integrating (\ref{phi0eq}) yields
\begin{eqnarray}
\chi_0=\phi_0-2 i c_1, \\
(r^2-c_2)^2-(\phi_0-i c_1)^2=c_1^2+c_2^2,
\end{eqnarray}
$c_1,c_2$ real constants with $c_1 \neq 0$ and $|r^2-c_2| < \sqrt{c_1^2+c_2^2}$ in order to have $\phi, \phi'$ imaginary.
By (\ref{eqthb},\ref{I1b}) one has
\be
\mathrm{d}b= i(\frac{\tau}{m}\w{\omega}^1-\frac{m}{\tau}\w{\omega}^2) B.
\ee
and hence
\be
\fl \mathrm{d}r \wedge \mathrm{d}m \wedge \mathrm{d}b = 2 i B Q \frac{m (\phi_0^2-\chi_0^2)}{r}(\frac{\tau}{m}\w{\omega}^3\wedge \w{\omega}^4 \wedge \w{\omega}^1-\frac{m}{\tau}\w{\omega}^3\wedge \w{\omega}^4 \wedge \w{\omega}^2)\neq 0,
\ee
showing that one can use $r,m$ and $b$ as coordinates. As $B$ satisfies the same equations (\ref{eqthb},\ref{I1b}) as $b$, it follows that $B=B(b)$, allowing one to redefine $b$ such that $B=1$. Defining a fourth coordinate $x$ by
\be
\frac{1}{2}(\frac{\tau}{m}\w{\omega}^1+\frac{m}{\tau}\w{\omega}^2)=h_1 \mathrm{d}b +h_2 \mathrm{d}r+h_3 \mathrm{d}m+\mathrm{d}x,
\ee
the equations (\ref{H1eq},\ref{H2eq}) imply
\begin{eqnarray}
\mathrm{d}H= \Sigma_1(b,m,r) \mathrm{d}b +(16 Qc_1mh_2+\Sigma_2(b,m,r)) \mathrm{d}r \nonumber \\
\ \ \ \ \ \ \ \ -(16Q c_1 m h_3-\frac{H-4Q(c_1+i\phi_0)}{m}) \mathrm{d}m -16Q c_1 m \mathrm{d}x, \label{Hvgl1}
\end{eqnarray}
showing that
\be
H=-16Q c_1 m x+2iQ(\phi_0+\chi_0)+m H_0,
\ee
with $H_0=H_0(b,r,m)$. It also follows that $h_2$ is a freely specifiable function of $b,r$ and $m$ (using a translation of $x$). Choosing $h_2=-mr / [4 Q c_1(c_1+i\phi_0)]$, (\ref{Hvgl1}) shows that $H_
0=H_0(b,m)$, such that $h_3$ can be made 0 by a further $(b,m)$-dependent translation of $x$. Herewith (\ref{Hvgl1}) eventually becomes
\be
\mathrm{d}H_0+u_0 \mathrm{d}b = 0
\ee
with
\be
u_0=16Q c_1 h_1-4 u -\frac{1}{2} H_0^2+16 Q c_1 x H_0+6m^2-128 c_1^2 x^2+12 Q r^2,\label{h1eq}
\ee
showing that $H_0=H_0(b)$ and $u_0=u_0(b)$. It remains to determine the function $E$ (or $u$) from (\ref{Eeq1},\ref{Eeq}) in order to obtain $h_1$ from (\ref{h1eq}). One readily finds
\be
h_1=\frac{xH_0}{2}-Q(4 c_1 x^2+\frac{m^2}{4 c_1})+F,
\ee
where $F=F(b)$ can be made 0 by a $b$-dependent translation of $x$ and a corresponding re-definition of $H_0$.
This gives the following expression for the dual basis ($\rho$, $\phi_0$ and $\chi_0$ being imaginary):
\begin{eqnarray}
\fl \frac{\tau}{m}\w{\omega}^1= \mathrm{d}x+\frac{iQmr}{4c_1(\phi_0-i c_1)}\mathrm{d}r+(\frac{H_0x}{2}-4Qc_1 x^2-\frac{2i c_1+Q m^2}{4c_1} )\mathrm{d}b ,\\
\fl \frac{1}{\rho} \w{\omega}^3= \frac{iQ}{4 c_1}(\frac{\mathrm{d}m}{m}-\frac{\phi_0}{\phi_0-ic_1}\frac{\mathrm{d}r}{r}) +i(Q\frac{r^2H_0-4m i\phi_0}{8r^2 c_1}-2  x) \mathrm{d}b,\\
\fl \rho \w{\omega}^4 = \frac{ir^2}{4 c_1}(-\frac{\mathrm{d}m}{m}-\frac{\chi_0}{\chi_0+ic_1}\frac{\mathrm{d}r}{r}) -i[\frac{r^2H_0-4mi\phi_0}{8c_1}-m-2Q r^2 x] \mathrm{d}b,
\end{eqnarray}
implying
\begin{eqnarray}
\fl R =-12 (Qr^2+m^2+E),\\
\fl E =-2m^2-\frac{8}{3}Q r^2+32c_1^2x^2+\frac{1}{12}H_0^2-\frac{1}{6}\frac{d H_0}{db}-4 Q c_1 x H_0-\frac{4Q}{3r^2}\phi_0\chi_0,\\
\fl H = (H_0-16 Q c_1 x)m+2iQ(\phi_0+\chi_0),\\
\fl \Phi_{00}=-\rho^2,\\
\fl \Phi_{22}=-\frac{r^4}{\rho^2},\\
\fl \Phi_{11}=-\frac{9}{2}Q r^2-\frac{7}{2}m^2-\frac{3}{2}E -\frac{4Q\phi_0\chi_0}{r^2},\\
\fl \Phi_{01}=\frac{\rho\tau}{r^2}(2\phi_0-3r^2),\\
\fl \Phi_{12}=-\frac{Q\tau}{\rho}(2\chi_0+3r^2),\\
\fl \Phi_{02}=-\tau^2.
\end{eqnarray}

\section{$KS_2\setminus KS_1$}

The analysis of the space-times belonging to $KS_2\setminus KS_1$ proceeds along the same lines, but is greatly facilitated by noticing that the conditions (\ref{eqthb},\ref{eqeta}) are transformed into each other under the Sachs~\cite{GHP} asterisk operation $\tho^*=\et, \thd^* =-\etd, \et^* = -\tho,\etd^*=\thd$. The same of course holds for the pairs of integrability conditions (\ref{I1a},\ref{I1b}), (\ref{I2a},\ref{I2b}), (\ref{condKS1},\ref{condKS2}). This becomes obvious after writing down the Sachs-transforms of $\phi,\phibar$ and $\phi'$ (recall that complex conjugation and the Sachs-operation do not commute):
\be
\phi^*=\frac{\rho\rd}{|\tau|^2} \phi,\ \ \phi'^*=\frac{\rho\rd}{|\tau|^2} \phi',\ \
\phibar^*=\frac{\rho\rd}{|\tau|^2} \phidbar,\ \ \phidbar^*=\frac{\rho\rd}{|\tau|^2} \phibar .
\ee
Defining coordinates $r$ and $m$ as in (\ref{rmdef}) one now finds $\phi=\phi(m)$, with
\be
\phi+\phibar=2 c_1 \neq 0 \label{c_1cond2},
\ee
and
\be
(\phi-c_1)^2-(m^2-c_2)^2 = c_1^2-c_2^2.
\ee
Here $c_1,c_2$ are real constants and $|m^2-c_2| < \sqrt{c_2^2-c_1^2}$ in order to obtain (\ref{c_1cond2}). Without further ado the results are presented below ($\rho$ and $\phi-c_1$ being imaginary):
\begin{eqnarray}
\fl \tau\w{\omega}^1= \frac{m^2}{4 c_1}(\frac{\mathrm{d}r}{r}-\frac{\phi-2c_1}{\phi-c_1}\frac{\mathrm{d}m}{m}) +[-Q\frac{4i Q r\phi-m^2H_0}{8c_1}+i(r+2iQ m^2 x)] \mathrm{d}a ,\\
\fl \frac{r}{\rho} \w{\omega}^3 = -i Q \mathrm{d}x+\frac{mr}{4c_1(\phi- c_1)}\mathrm{d}m-i (\frac{H_0x}{2}-4c_1 x^2-\frac{r^2}{4c_1}+ Q)\mathrm{d}a ,\\
\fl \frac{\rho}{r} \w{\omega}^4 = i \mathrm{d}x-Q\frac{mr}{4c_1(\phi- c_1)}\mathrm{d}m+i (\frac{ H_0x}{2}-4c_1 x^2-\frac{r^2}{4c_1})\mathrm{d}a,
\end{eqnarray}
implying
\begin{eqnarray}
\fl E =-2Q r^2-\frac{8}{3}m^2+32Qc_1^2x^2+\frac{Q}{12}H_0^2-\frac{1}{6}\frac{d H_0}{da}-4 Q c_1 x H_0+\frac{4Q}{3m^2}\phi \overline{\phi},\\
\fl H = (H_0-16 c_1 x)r+4i(\phi-c_1),\\
\fl \Phi_{11}=\frac{9}{2}m^2+\frac{7}{2}Q r^2+\frac{3}{2}E -\frac{4\phi \overline{\phi}}{m^2},\\
\fl \Phi_{01}=\frac{\rho\tau}{m^2}(2\phi-3m^2),\\
\fl \Phi_{12}=-\frac{Qr^2\tau}{m^2\rho}(2\phi+3m^2),
\end{eqnarray}
with the expressions for $R,\Phi_{00},\Phi_{22}$ and $\Phi_{02}$ being identical to those of the previous paragraph.

As the Sachs transform destroys the reality conditions on the null-tetrad, it is to be expected that the two solution families will be related by a \emph{complex} coordinate transformation, in combination with a possible complex transformation of the parameters and the free functions. Indeed such a transformation exists and one can verify that the $KS_1$ line-elements are obtained by applying the following operations to $KS_2$:
\begin{eqnarray}
r \rightarrow m Q^{-1/2}, \ m \rightarrow r q^{1/2}, \ a \rightarrow b Q^{1/2}, \ x \rightarrow -i Q^{1/2} x + g, \\
c_1 \rightarrow i Q c_1,\ \phi \rightarrow q \phi_0, \ h_0 \rightarrow Q^{1/2} h_0 -16 q c_1 g,
\end{eqnarray}
with $g(b)$ a solution of
\be
2 g' + Q^{1/2} - i h_0 g +\frac{8 i c_1}{q} Q^{-1/2} g^2 = 0 .
\ee

\section{Summary}
All KS space-times are constructed in which one of the integrability conditions (\ref{condKS1},\ref{condKS2}) (both of which
 were assumed to be valid in \cite{Jeffryes1984}) are violated. This leads to the conformal classes of $KS_1$ and $KS_2$ space-times
discussed in paragraphs 3 and 4. Each of these classes is characterized by a single arbitrary function ($H_0$). In general none of the
corresponding space-times admits any isometries (when $H_0$ is constant then a one-dimensional group of isometries exists), but has a conformal representant $(\mathcal{M},\hat{g})$, admitting a Killing tensor
with precisely one constant eigenvalue. The explicit form of the resulting space-times raises doubts about the correctness of the
Hauser-Malhiot metrics \cite{HauserMalhiot1978} admitting a conformal Killing tensor with one constant eigenvalue. It is an open
problem whether the classes $KS_1$ and $KS_2$, together with the metrics presented in \cite{Jeffryes1984} and \cite{McLenVdB1993}
exhaust the full set of KS space-times. There is also no guarantee that the classes $KS_1$ and $KS_2$ discussed in the present paper
 have a conformal representant possessing any physically interesting interpretation. These issues will be dealt with in a forthcoming
publication.

\section{Acknowledgment}
I thank Brian Edgar (Link\"oping University) for his comments on an earlier version of this paper.

\section*{References}


\providecommand{\newblock}{}

\end{document}